\documentclass{elsart}
\usepackage{graphicx,amssymb}
\journal{Chemical Physics}

\newcommand{\AmS}{{\protect\the\textfont2
  A\kern-.1667em\lower.5ex\hbox{M}\kern-.125emS}}

\begin{document}
\begin{frontmatter}
\title{Decoherence and dephasing in  coupled Josephson-junction qubits}
\author[FTP,MICH]{Michele Governale},
\author[MIL]{Milena Grifoni}, 
\author[FTP]{Gerd Sch\"on}
 \address[FTP]{Institut f\"ur Theoretische Festk\"orperphysik
Universit\"at Karlsruhe,
Engesser Stra\ss e 6, D-76128 Karlsruhe}%
\address[MICH]{ Dipartimento di Ingegneria dell'Informazione, via Diotisalvi 2, I-56126, Pisa, Italy} 
\address[MIL]{Departement of Applied Sciences, Delft University of Technology, Lorentzweg 1,  2628 CJ Delft, The Netherlands}

\begin{abstract}
 We investigate the decoherence and dephasing of two coupled Josephson qubits.
 With  the 
 interaction between the qubits being generated by current-current 
 correlations,  two different situations in which the qubits are coupled 
 to the same bath,
 or to two independent baths, are considered. Upon focussing on 
 dissipation being caused by the fluctuations of voltage sources,
 the relaxation and dephasing rates are explicitly evaluated.
 Analytical and numerical results for the coupled qubits dynamics are 
 provided.  
\end{abstract}

\end{frontmatter}
\noindent

\section{Introduction}

Nano-electronic devices have recently attracted much interest
as possible candidates for the implementation of quantum computers.
In fact, in contrast to ion traps   \cite{ions} and NMR systems \cite{NMR},
nano-electronic devices
 are  easily embedded in an electronic circuit, and  
can be scaled up to large number of quantum bits (qubits) and quantum gates.   
Ultra small quantum dots with discrete levels, and in particular spin degrees of freedom embedded in nano-scale structured materials have been proposed \cite{qdots}. However, 
these systems are difficult to fabricate in a controlled way. More suitable candidates 
are Josephson contact systems, where the coherence of the superconducting state can be
exploited, and which can be fabricated by well established litographic methods. 
Macroscopic  flux states in a SQUID have been proposed as basic quantum bits \cite{Mooij}.
Spectroscopy experiments of the flux qubit spectrum have recently been performed \cite{Caspar}.   

In this work we consider the alternative design which uses the charge in low-capacitance junctions as quantum degree of freedom \cite{RMP,Shnirman}. The logical states of the qubit are 
then macroscopic states differing by one Cooper pair charge.  
In a recent experiment Nakamura et al. have observed time-resolved coherent
oscillations in such a Josephson junction set-up \cite{Nakamura}.
Several possible realizations have been by now discussed \cite{Shnirman,Averin,Makhlin}.
 The set up we consider is discussed in Section 2, and it follows the
design proposed in \cite{Makhlin}. In particular, single bit and two-bit operations may
be performed by applying a sequence of suitably tailored gate voltages.
 
Quantum computation requires a phase coherent time evolution. Hence, it is of prominent 
importance to deal with systems possessing long intrinsic phase coherence times, 
and to minimize possible external sources for dephasing \cite{qtheory}. 
Moreover, crucial for quantum computation is
 the ability to perform calculations  with single as well as with  coupled qubits.
Previous works analyzed only dephasing times of single qubits \cite{Weiss99}, or gave an estimate for $N$ uncorrelated qubits \cite{RMP}.
Only a few previous works investigated the dynamics of coupled spins embedded in a thermal environment \cite{Stamp}.
 In this work we analyze the 
influence of dissipation and of thermal quantum fluctuations on the coherent evolution of  
two coupled qubits. We distinguish between two configurations: In the first one
 the two qubits are dephased by the same thermal bath. In the second one
each qubit is coupled to its own environment. 
The general characteristics  will be the theme of Section 3, where  the phase coherence time and
the decoherence rate will be explicitly evaluated for the two different 
environmental configurations.  In Section 4   we provide
 analytical and numerical results for the population dynamics,
 when dissipation is dominated by fluctuations of the voltage sources.  
Finally, in Section 5 we  draw some conclusions. 

\section{Josephson Junction Qubits}


\subsection{The Josephson-junction  qubit} 

In the following we briefly discuss the  design for tunable  Josephson charge qubits following \cite{Shnirman,Makhlin}. It allows to manipulate the qubits parameters independently. The system  consists of a tunable Cooper pair box,  where  a small superconducting island is connected to two  tunnel junctions embedded in the two arms of a SQUID loop threaded by a flux $\Phi_X$, cf. Fig. 1. The junctions are supposed to be identical and are characterized by a  capacitance $C^0_J$ and coupling energy $E_J^0$.  In this situation, an ideal voltage source 
 $V_g$ is connected to the system via a gate capacitor $C_g$.
If the self-inductance of the SQUID is low, the two junctions effectively behave
as a single junction being characterized by the effective 
 capacitance $C_J=2C^0_J$, and by the tunable Josephson energy 
\begin{equation}
E_J(\Phi_X)=2E_J^0\cos(\pi\Phi_X/\Phi_0)\;,
\end{equation} 
 where $\Phi_0=h/2e$ is the flux quantum.
 Then, the charging energy of the 
 superconducting island is characterized by the scale $E_{C}=e^2/(2C_g+2C_J)$. 
In the following we consider the situation where the superconducting energy gap
 is the largest energy in the problem. Then {\em only} Cooper pairs tunnel through the 
superconducting junctions.   In the regime $E_J\ll E_{C}$ a convenient basis is formed by the charge states parametrizing the number $n$ of Cooper pairs on the island.   
In this situation the  dimensionless gate charge $n_g=C_gV_g/(2e)$ acts as a control field. 
For values of  $n_g$  close to a half integer two adjacent Cooper pair states become
close to each other. We concentrate on the voltage interval near a degeneracy point where
 two charge states, e.g. $n=0$ and $n=1$ play a role, while higher charge states, having
 much higher energy can be ignored. Hence, in the absence of dissipation, the system effectively reduces to a two-state system being described by the Hamiltonian
\begin{equation}
H_{\rm qb}=-\frac{1}{2}E_z(n_g)\sigma_z-\frac{1}{2}E_J(\Phi_X)\sigma_x\;,
\label{Hqbit}
\end{equation} 
where the $\sigma$'s are Pauli spin matrices, and $E_z=4E_{C}(2n_g-1)$ is controlled by
the gate voltage. In this representation the charge states $n=0$ and $n=1$ correspond to
 the vectors $|\downarrow\rangle$ 
 and $|\uparrow\rangle $ 
respectively. They are eigenstates of $\sigma_z$ with eigenvalues 
$-1$ and $1$, respectively. Hence, for a qubit being prepared in one of 
the eigenstates of $\sigma_z$, quantum coherent oscillations with frequency
$E_{\rm qb}/\hbar$, with 
\begin{equation}
E_{\rm qb}:=\sqrt{E_z^2+E_J^2}\;,
\label{Eqb}
\end{equation}
occur.

\subsection{The electromagnetic environment}

An ideal quantum system preserves quantum coherence, i.e., its time evolution 
is determined by deterministic { reversible} unitary transformations.
In reality, any  physical quantum system is subject to various disturbing factors which
act in destroying the  phase coherence.   
 The concept of quantum computation heavily relies on the possibility of 
realizing quasi-ideal systems. Hence, it is crucial to investigate the effects of the
environment on the qubits, and to understand how the environment induced dephasing can
be {minimized}. It should also be realized that coherent quantum manipulations of the qubits are
still possible if the dephasing time is finite but not too short. In fact, the quantum
 error-correction techniques \cite{Steane} allow to correct errors if they do not occur
too often.  

In a charge Josephson qubit, the system is sensitive to the electromagnetic
fluctuations in the external circuit and in the substrate, and to background charge 
fluctuations. We  start by considering here the dissipative effects which arise  from the 
fluctuations of the voltage sources. The equivalent circuit of a 
qubit coupled to an impedence $Z(\omega)$ is shown in Fig. 1. 
When $Z(\omega)$ is embedded into the circuit (with $E_J=0$), the voltage fluctuations
between the terminals of $Z(\omega)$ are characterized by the equilibrium  
correlation function \cite{RMP,Weiss99}
\begin{eqnarray}
\lefteqn{\langle \delta V(t) \delta V(0) \rangle_{}}\nonumber\\ &=&
\frac{1}{\pi}\int_0^\infty d\omega  {\rm Re}\{Z_t(\omega)\} \hbar\omega
   \left(\coth
\left(\frac{\hbar\omega}{2k_{\rm B}T}\right)\cos(\omega t)-i\sin (\omega t)\right)\;.
\label{fluctuations}
\end{eqnarray}
Here $Z_t(\omega):=[i\omega C_t + Z^{-1}(\omega)]^{-1}$,
 with $C_t^{-1}=C_J^{-1}+C_g^{-1}$, 
 is the total impedence between the
terminals. 
Following \cite{Caldeira} we model the dissipative influence of $Z_t(\omega)$ as resulting
from a bath of harmonic oscillators  described by the Hamiltonian
\begin{equation}
H_{\rm B}=\sum_i \left[ \frac{p^2_i}{2m_i} +\frac{m_i\omega^2_ix_i^2}{2}\right]\;.
\end{equation}
It is assumed that the voltage between the terminals of $Z_t(\omega)$ is given by
 $\delta V =\sum_i\lambda_i x_i$, and the
 spectral density $G(\omega):=\frac{\pi}{2}\sum_i\frac{\lambda_i^2}{m_i\omega_i}\delta (\omega -\omega_i)$ is
 chosen to reproduce the fluctuations spectrum (\ref{fluctuations}), i.e., 
 $
 G(\omega)=\omega {\rm Re}Z_t(\omega)$.
 Thus, embedding the element $Z(\omega)$ in the qubit circuit one arrives, 
in the two-state approximation, to the spin-boson Hamiltonian \cite{RMP}
\begin{equation}
H_{\rm SB}= H_{\rm qb}+ H_{\rm B}+\frac{1}{2}\sigma_z X \;,
\label{spin-boson}
\end{equation} 
where $X:=2e (C_t/C_J)\delta V$ describes the coupling to the bath. 
 In the following we concentrate on the fluctuations due to an Ohmic resistor $Z(\omega)=R$. It is  convenient to introduce the 
 spectrum $J_V(\omega):= 4e^2(C_t/C_J)^2 G(\omega)$. 
For an Ohmic resistor it assumes the form 
$J_V(\omega)=2\pi\hbar\alpha\omega$ which is linear 
at low frequencies up to some cut-off $\omega_c=(RC_t)^{-1}$. 
 The dimensionless parameter $\alpha$
 which characterizes the strength of the dissipative effects reads
\begin{equation}
\alpha=\frac{4R}{R_{\rm K}}\left(\frac{C_t}{C_J}\right)^2\;,
\label{alpha}
\end{equation}
where $R_{\rm K}=h/e^2\approx 25.8$k$\Omega$ is the quantum resistance.
Hence, in order to make dissipation small one has to use a voltage source with very low 
resistance, and choose the gate capacitance $C_g\approx C_t\ll C_J$ as small as possible.
For a typical resistance $R\approx 50\Omega$ one has $\alpha \approx 10^{-2}(C_g/C_J)^2$.
Upon choosing $C_J\approx 100 aF$ and $C_g\approx 1aF$ one can reach coupling parameters
as small as $\alpha\approx 10^{-6}$.

A similar line of reasoning can be performed to include  also  the effect of $1/f$ noise due to background charges \cite{Paladino}. It is found that the effect of the background charges can be mapped into that of a harmonic oscillator bath with a spectrum that may have a simple behavior  $J_{1/f}(\omega)\propto K/\omega$. Usually, such fluctuations occur on a longer  time scale than the voltage fluctuations. They also limit the coherence of the system. However in our
 charge qubit set-up dissipation is dominated by voltage flucutuations.
 In the following we shall mainly consider the effects of a harmonic bath being
characterized by the Ohmic spectral density $J_V(\omega)$.
However, how to include $1/f$ noise  in our formalism will be discussed in the next Section . 

Finally, one can consider the effects of the fluctuations
of the externally supplied flux through the SQUID loop of the charge qubit. These fluctuations couple to the qubit variable $\sigma_x$. We relate them to an effective
 impedence $R_I$, assumed to be real, of the current circuit. At typical high frequencies
 of the qubit's operation this resistance is of the order of the vacuum impedence
 $R_I\approx 100\Omega$. This yields for the dimensionless coupling constant \cite{RMP}
\begin{equation}
\alpha_x=\frac{R_{\rm K}}{R_I}\left(\frac{M}{\Phi_0}
\frac{\partial E_J(\Phi_X)}{\partial \Phi_X}\right)^2\;,
\label{alphax}
\end{equation}
with $M$ being the mutual inductance. For $M\approx 0.01$nH and $E_J^0\approx0.1$K one obtains $\alpha_x\approx 10^{-8}$.
Thus, these latter fluctuations seem not to be dominant, and will be neglected in the following treatment. They can however be easily included in our formalism, as we shall discuss in Section 3.

\subsection{Coupled qubits}

For quantum computation  pairs of qubits have to be coupled in a controlled way.
In the scheme proposed in Fig. 2 all qubits are coupled by one mutual inductor $L$. For $L=0$ the system reduces to a series of uncoupled qubits, 
while they are coupled strongly
when $L\to\infty$.
Hence, for finite $L$, the total Hamiltonian of this system consists
 of $N$ qubits contributions as in (\ref{Hqbit}), and of an oscillator resulting from the
 inductance and the total capacitance $NC_t$, with $C_t^{-1}=C_J^{-1}+C_g^{-1}$, of all 
qubits. The voltage oscillations in the $LC_t$ circuit affect all the qubits equally. They 
 induce  a phase shift $2\pi\varphi=2\pi\frac{C_t}{C_J}\frac{\Phi}{\Phi_0}$ of the phase variables $\Theta_i$ conjugated to the Cooper pair number $n_i$ of the $i$-th qubit.
Here $\Phi$ denotes the flux in the mutual inductor, and $\Phi_0=h/2e$ is the flux quantum.
In the following the parameters are chosen such that  the oscillator remains in the
 ground state for all relevant operation frequencies, i.e.,
$E_C,E_J\ll \hbar\omega_{L}$, where $\omega_L=(NLC_t)^{-1/2}$ is the characterisitic oscillator frequency. Moreover, we  assume that fluctuations of the phase shift $\varphi$ are small
compared to unit, i.e., $\frac{C_t}{C_J}\sqrt{\langle \Phi^2\rangle }\ll\Phi_0$.
 Since $\langle \Phi^2\rangle/L \approx \hbar\omega_L/2  $ this condition imposes only
 weak constraint on the parameters. Then, although the $ LC_t$ oscillator remains in its
 ground state, it provides an effective coupling between the qubits. It reads 
\cite{Makhlin}
\begin{equation}
H_{\rm int}=-\sum_{i<j}\frac{E_J(\Phi_{Xi})E_J(\Phi_{Xj})}{E_L}\sigma_y^{(i)}\sigma_y^{(j)}\;,
\end{equation} 
with the energy scale
\begin{equation}
E_L=\left(\frac{C_J}{C_g}\right)^2\frac{\Phi_0^2}{\pi^2L}\;,
\end{equation}
while $E_J^i=E_J(\Phi_{Xi})$ are the effective Josephson energies of the qubits,
 controlled by the external flux. 
This coupling energy can be easily understood as the magnetic energy of the current
in the inductor, where the current is the sum of the contributions from the qubits with non
 zero Josephson coupling, $I_i\propto E_J^i\sigma_y^{(i)}$. Noticeably the strength of the 
interaction does not depend on the number of qubits in the system. However, the frequency 
$\omega_L$ of the oscillator scales with $1/\sqrt{ N}$. This limits the allowed 
number of qubits in the system, since this frequency should not drop below typical
 eigenfrequencies of the qubit. The interaction scale $E_L$ involves the screening ratio
 $C_g/C_J$. As discussed in the preceding subsection, this ratio should be taken as small
as possible to minimize the decoherence effects of the dissipative electromagnetic environment. Consequently, to achieve a reasonable interaction strenght a large inductance is needed.  
For example, for the typical values $E_J\approx 100 mK$ and $C_g/C_J\approx 0.1$ an 
inductance $L\approx 1\mu$H is needed. A more sophisticated set-up which overcomes this 
problem enabling to use smaller inductances is discussed in \cite{RMP}.

 From the above discussion it is clear that, with tunable Josephson couplings $E_J(\Phi_X)$, 
  two-qubits gate operations can be performed by setting to zero all the 
 couplings except for two selected qubits, say 1 and 2. 
The resulting two-qubits Hamiltonian takes the form:
\begin{eqnarray}
H_{2\rm qb}=-\frac{E^1_J}{2}\sigma_x^{(1)}-\frac{E^2_J}{2}\sigma^{(2)}_x 
-\frac{E^1_z}{2}\sigma_z^{(1)}-\frac{E^2_z}{2}\sigma_z^{(2)} -
\frac{E^{1}_JE^2_J}{E_L}\sigma_y^{(1)}\sigma_y^{(2)}\;.
\label{H2qbit}
\end{eqnarray} 
Finally, to include the environmental influence, we distinguish between the two
different situations shown in Figs. 2a and 2b. In case I, shown in Fig. 2a, all
 the qubits are coupled to the same oscillator bath. The resulting Hamiltonian then reads:
\begin{equation}H_{\rm I}=H_{\rm 2qb}+ H_{\rm B}+\frac{1}{2}(\sigma_z^{(1)}+\sigma_z^{(2)})X\;,
\label{HI}
\end{equation}
with $X$ defined below (\ref{spin-boson}). As we shall see in the next section, this case reveals interesting symmetries.
In case II, shown  in Fig. 2b, each qubit is coupled to its own  
oscillator bath.  It is described by 
the Hamiltonian
\begin{equation}
H_{\rm II}=\sum_{i=1,2}H_{\rm SB}^i  +\frac{\gamma}{2}\sigma_y^{(1)}\sigma_y^{(2)}\;,
\label{HII}
\end{equation}
with $H_{\rm SB}$ being the spin-boson Hamiltonian introduced in (\ref{spin-boson}), and where the coupling parameter
\begin{equation}
\gamma=
- 2\frac{E^{1}_JE^2_J}{E_L}
\end{equation}
was introduced. This  case of {\em uncorrelated} baths is thought to better describe the situation in charge Josephson qubits. 

\section{Dissipation and dephasing of coupled qubits}

In this section we explicitly solve the dynamical problem posed by the
dissipative two qubit Hamiltonians (\ref{HI}) and (\ref{HII}). 
To better outline the symmetries and possible  constants of motion of the problem,  we find it convenient to  discuss the 
coupled qubit evolution in the Hilbert space spanned by the triplet  states 
 $|\uparrow \uparrow \rangle:=(1,0,0,0)^T$, $(|\uparrow \downarrow \rangle+|\downarrow \uparrow \rangle)/\sqrt{2}:=(0,1,0,0)^T$, 
  together with $ |\downarrow \downarrow \rangle:=(0,0,1,0)^T$, as well as by 
  the singlet state 
 $(|\uparrow \downarrow \rangle-|\downarrow \uparrow \rangle)/\sqrt{ 2}
:=(0,0,0,1)^T$. 
In this representation the dissipative Hamiltonian 
 for  uncorrelated baths assumes the appealing form 
\begin{eqnarray}
 H_{\rm II} = 
 \begin{array}{rcl}
 -\frac{1}{2}
 {\displaystyle 
\left( \begin{array}{cccc}  \varepsilon - s & \eta & \gamma & 
 -\Delta\eta \\[1mm]
 \eta  &   -\gamma & \eta &
 -\Delta\varepsilon + \Delta s\\[1mm]
 \gamma       & \eta  & -\varepsilon +s  & \Delta \eta \\[1mm]
  -\Delta\eta & -\Delta\varepsilon + \Delta s & \Delta \eta  & \gamma 
\end{array}
 \right)  }
\end{array}+ H_{\rm B}\;. 
\label{Hsigmazqbit}
\end{eqnarray}
Here, for simplicity of notation, we introduced the quantities 
\begin{eqnarray}
\Delta\varepsilon & :=&E_z^1-E_z^2\;,\;\;
\Delta\eta:=(E_J^1-E_J^2)/\sqrt{2}\;, \nonumber\\
\varepsilon &:=& E_z^1+E_z^2\;,\;\;\eta := (E_J^1+E_J^2)/\sqrt{2}\;,
\end{eqnarray}
characterizing the isolated coupled qubits. The  sum and difference 
\begin{equation}
s := X_1+X_2 \;,\qquad \Delta s =X_1-X_2\;,
\end{equation}
characterize  the baths coupling, cf. below (\ref{spin-boson}) for the definition  $X_i$ for the single qubit $i$.  

For the Hamiltonian $H_{\rm I}$ of qubits coupled to a single bath 
the same form as for $H_{\rm II}$ applies upon substitution of 
\begin{equation}
 s\to 2X \;,\qquad \Delta s\equiv 0\;.
\label{subst}
\end{equation} 

\subsection{The degenerate case}

In the following we focus on the 
 the {\em degenerate case}
$\Delta\eta=\Delta\varepsilon=0$, corresponding to equal Josephson energies and asymmetries of the two qubits. 
 It is in this situation that quantum effects arising from the coupling 
between the qubits are expected to play a major role. 
Then the  
{\em non}dissipative Hamiltonian 
$H_{\rm 2qb}$ becomes separated into two blocks, corresponding to the triplet 
and singlet states, respectively.
To be definite, the  unperturbed Hamiltonian  satisfies the Schr\"odinger equation
$H_{\rm 2qb}|n\rangle=E_n|n\rangle$, with eigenvalues
\begin{eqnarray}
E_{1,4}=\mp \frac{1}{2}\sqrt{\varepsilon^2 + \gamma^2 + 2\eta^2 }:=\pm 
\frac{E}{2}\;,\qquad
E_{2,3}=\mp\frac{\gamma}{2}\;,
\end{eqnarray} 
and eigenfunctions, expressed in the singlet/triplet basis, 
\begin{eqnarray}
|1\rangle&=& \frac{1}{2}\frac{1}{\sqrt{E(E+\gamma)}}
(E+\gamma+\varepsilon,2\eta,E-\varepsilon +\gamma,0)^T\;,
\nonumber \\
 |2\rangle&=&(0,0,0,1)^T\;,\qquad 
|3\rangle =\frac{1}{\sqrt{2\eta^2+\varepsilon^2}}(\eta,-\varepsilon,-\eta,0)^T \;\nonumber\;, \\
|4\rangle&=& \frac{1}{2}\frac{1}{\sqrt{E(E-\gamma)}}
(E-\varepsilon -\gamma,-2\eta,E+\varepsilon-\gamma,0)^T\;.
\label{transform}
\end{eqnarray}
Hence, if the system was prepared in one of the triplet states, it will undergo
quantum coherent oscillations among  the three states with oscillation frequencies $\omega_{m n}:=|E_m-E_n|/\hbar$, with $m,n = 1,3,4$. However
 {\em no} motion 
is associated with the singlet state, being the eigenstate $|2\rangle$ of
the unperturbed Hamiltonian. 
We observe that the maximal oscillation frequency is 
 $\omega_{41}= E/\hbar > \sqrt{2}E_{\rm qb}/\hbar$, where $E_{\rm qb}$ is the single
 qubit tunneling splitting (\ref{Eqb}).
In the case of distinct baths, 
 dissipation mixes the triplet and singlet states. To be definite, the total Hamiltonian for {\em un}correlated baths reads $H_{\rm II}=H_{\rm II}^0+H_{\rm B}$ with
\begin{eqnarray} 
\begin{array}{rcl}
H_{\rm II}^0 &=& -\frac{1}{2}
 {\displaystyle
\left( \begin{array}{cccc}  \varepsilon -s  & \eta & \gamma & 
 0 \\[1mm]
 \eta  &  -\gamma  & \eta &
 \Delta s\\[1mm]
 \gamma       & \eta  & -\varepsilon + s & 0 \\[1mm]
  0 & \Delta s& 0 & \gamma 
\end{array}
  \right)  }
\end{array}. 
\end{eqnarray}
Because the problem of coupled qubits 
 is isomorph to that of two interacting spins in the presence
of dissipation, it is suggestive to associate to the triplet and singlet states
  quantum numbers $(S,S_z)$, being eigenvalues of the  squared total spin operator  $S^2$,
 and of the projection $S_z$ of the total spin  on the $z$-axis. Then, the triplet states are
 characterized by $|S=1,S_z=-1,0,1\rangle$, while the singlet state is given by $|S=0,S_z=0\rangle$. 
Hence, once the system is prepared, e.g., in the entangled state with quantum numbers $|S=0,S_z=0\rangle$, it will stay in that state for ever if no dissipation is present. In fact, as appearent from 
 inspection of $H_{\rm II}^0$, the effect of dissipation is to induce transitions
between the states $|S=0,S_z=0\rangle$ and $|S=1,S_z=0\rangle$, yielding thermalization also for the singlet state. 

Thermalization of the singlet state however is  impeded in the case of  dissipative qubits   being characterized by the Hamiltonian $H_{\rm I}=H_{\rm I}^0+H_{\rm B}$.
In fact, in the degenerate case $H_{\rm I}^0$ is obtained from  $H_{\rm II}^0$
upon performing the substitution (\ref{subst}).
When dissipation is added,  the oscillation 
 frequencies $\omega_{nm}$, $m \neq n$, acquire a finite dephasing rate  $\Gamma_{mn}$.  Moreover, 
 the system relaxes to thermal equilibrium with a decay rate 
 $\Gamma$
 being associated with incoherent tunneling processes. 
These rates are explicitly evaluated in the next section. 

\subsection{Redfield equations}

To perform quantitative calculations we apply the well established 
Bloch-Redfield formalism \cite{Redfield,Argyres} 
to the Hamiltonians $H_{\rm I}$ and $H_{\rm II}$.
 The Redfield approach 
 provides a set of coupled master-equations for the matrix elements of the
reduced density matrix. 
Upon performing a Markov approximation, and in the basis
of the eigenstates of  the unperturbed Hamiltonian, they read
\begin{equation}
\label{block}
\dot{\rho}_{nm}+i \omega_{nm}\rho_{nm}=\sum_{n\prime,m\prime} 
{\mathbf R}_{n,m,n\prime,m\prime}\rho_{n\prime m\prime}\;,
\label{Redfieldeq}
\end{equation}
where  $\mathbf{R}$ is the 
 Redifield tensor whose  elements    are given by
\begin{eqnarray}
\!\!\!\!\!\!\!
{\mathbf R}_{n,m,n\prime,m\prime}={\mathbf \Lambda}_{m\prime,m,n,n\prime}+
{\mathbf \Lambda}^*_{n\prime,n,m,
m\prime} -
\sum_k\left[{\mathbf \Lambda}_{n,k,k,n\prime}\delta_{m,m\prime}+
{\mathbf \Lambda}^*_{m,k,k,m\prime}\delta_{n,n\prime}\right].
\end{eqnarray}
For the Hamiltonian $H_{\rm II}$
 the tensor ${\mathbf \Lambda}$  assumes the form 
\begin{eqnarray}
{\mathbf \Lambda}^{\rm II}_{n,m,n\prime,m\prime}&=&\frac{1 }{\hbar^2}(s_{nm}s_{n\prime m\prime} +r_{nm} r_{n\prime m\prime}  ) \hat M_{\rm II}(\omega_{n\prime m\prime}) 
\nonumber \\
&-&\frac{1 }{\hbar^2}(s_{nm} r_{n\prime m\prime} + r_{nm} s_{n\prime m\prime}) \Delta \hat M(\omega_{n\prime m\prime})\;,
\end{eqnarray}
where, $\hat M_{\rm II}(\Omega ):=\hat M_1(\Omega)+\hat M_2(\Omega)$,
$\Delta \hat M(\Omega):=\hat M_1(\Omega)-\hat M_2(\Omega) $, with
$\hat M_i$, ($i=1,2$), reading 
\begin{equation}
M_i =\frac{1}{4}\int_0^\infty dt e^{-i \Omega t}
\langle X_i(t)X_i(0) \rangle \;.
\end{equation}
Upon introducing the spectral densities 
$J_{\rm II}(\omega):=[J_{V1}(\omega)+J_{V2}(\omega)]/4$ and $\Delta J(\omega):=[J_{V1}(\omega)-J_{V2}(\omega)]/4$, cf. below (\ref{spin-boson})
 for the definition of $J_V$ for the single qubit, 
 the one-sided Fourier transforms can be explicitly evaluated from (\ref{fluctuations}).  One finds 
\begin{eqnarray}
{\rm Im} \hat M_{\rm II}(\Omega)&=& \frac{\hbar}{\pi}
\int_0^{\omega_c} d\omega \frac{J_{\rm II}(\omega)}{\omega^2-\Omega^2}
\left[\Omega\coth\left( \frac{\hbar\omega}{2k_{\rm B}T}\right)-\omega \right]
\;,\nonumber \\ 
{\rm Re} \hat M_{\rm II}(\Omega)&=& \frac{\hbar}{2}J_{\rm II}(\Omega)\left[
\coth \left( \frac{\hbar\Omega}{2k_{\rm B}T}\right)-1\right]\;,
\label{Mom}
\end{eqnarray}
where $\omega_c$ is the characteristic cut-off of the bath spectral density
 $J_{\rm II}$, cf. below Eq. (\ref{spin-boson}).
Likeways $\Delta \hat M (\omega)$  has the same form as (\ref{Mom}) but with
$J_{\rm II}(\omega)\to \Delta J_{} (\omega)$.
 In the following, for simplicity, we focus on the case of equal baths,
 yielding $\Delta \hat M\equiv 0$. 
  Finally,  $s_{mn}$ and $r_{nm}$ are the 
  matrix elements of the operators  
$\hat s=(\sigma_z^{(1)} +  \sigma_z^{(2)})/2$ and 
$\hat r= (\sigma_z^{(1)} - \sigma_z^{(2)})/2$. We note that 
 $s_{nm}$ derives from transitions within the triplet space, while
$r_{nm}$ describes transitions between the singlet and the triplet
state with the same quantum number $S_z=0$. 
To be definite these matrix elements read:
\begin{eqnarray}
s_{11}&=&-s_{44}=\frac{\varepsilon}{E}\;,\qquad s_{22}=s_{33}=0\;,\qquad
s_{13}=\frac{\eta\sqrt{E+\gamma}}{\sqrt{E(2\eta^2+\varepsilon^2)}}
\nonumber\\ 
s_{34}&=&\frac{\eta\sqrt{E-\gamma}}{\sqrt{E(2\eta^2+\varepsilon^2)}}\;,\qquad 
s_{14}=-\frac{\gamma\varepsilon}{E\sqrt{(2\eta^2+\varepsilon^2)}}\;,
\end{eqnarray}
$s_{nm}=s_{mn}$ and  $s_{2j}=s_{j2}=0$ for any $j$. Finally, the only non vanishing matrix elements $r_{nm}=r_{mn}$ are 
\begin{eqnarray}\!\!\!\!\!\!\!\!\!\! 
r_{12}=\frac{\eta}{\sqrt{E(E+\gamma)}}\;, \qquad
r_{24}=-\frac{\eta}{\sqrt{E(E-\gamma)}}\;,\qquad 
r_{23}=-\frac{\varepsilon}{\sqrt{(2\eta^2+\varepsilon^2)}}.
\end{eqnarray}
For the Hamiltonian $H_{\rm I}$  the tensor ${\mathbf \Lambda}$   reads  
\begin{eqnarray}
{\mathbf \Lambda}^{\rm I}_{n,m,n\prime,m\prime}=\frac{1 }{\hbar^2}s_{nm}s_{n\prime m\prime}    \hat M_{\rm I}(\omega_{n\prime m\prime})
\end{eqnarray}
where,
\begin{eqnarray}
\hat M_{\rm I}(\Omega)&=&\int_0^\infty dt e^{-i \Omega t}
\langle X(t)X(0)\rangle \;.
\end{eqnarray}
Hence, $\hat M_{\rm I}(\Omega)$ has the same form as (\ref{Mom}) with
$J_{\rm II}(\omega)\to J_{\rm I}(\omega)\equiv J_V(\omega)$. 
Note that, for the case of equal baths here considered, 
\begin{equation}
J_{\rm I}(\omega)=2J_{\rm II}(\omega)=2\pi\hbar\alpha\omega\;,
\end{equation}
with $\alpha$ defined in (\ref{alpha}). Thus, apparently, dissipative effects
are {\em stronger} for the dissipative configuration I.
Some comments are necessary:\\
(i) These results are valid for a  dissipative coupling to the qubit variable $\sigma_z$. A generalization which includes dissipation parallel to the
other spin matrices $\sigma_i$, as for example the flucutuating contribution
(\ref{alphax}), is straightforward. In fact, 
such contributions simply modify additively  the tensor  
${\mathbf \Lambda}_{n,m,n\prime,m\prime}$. It acquires extra terms which are proportional to the matrix elements of the operators $\sigma_i^{(1)}\pm \sigma_i^{(2)}$, and of the spectral density of the new dissipative contribution.\\
(ii) As discussed in the previous section, together with voltage fluctuations
possessing an Ohmic spectrum, 
 the electromagnetic environment provides a 
 $1/f$ contribution arising from the background charges. 
This noise also couples to the $\sigma_z$ operator of the single qubit. 
Hence, to take into account this noise, as well as other noise 
sources coupling to $\sigma_z^{}$, the total spectral density
for the single qubit capturing the effects of $\sigma_z^{}$ coupling
 is modified to 
\begin{equation}
J_z(\omega)=J_{V}(\omega)+J_{1/f}(\omega)+...
\end{equation} 
This introduces different time scales for the decay of noise-induced
 correlations. In particular, those associated to $1/f$ noise decay on a very
long scale compared to the qubit dynamics. Hence, the Markov approximation
invoked to obtain the set of equations (\ref{Redfieldeq}) breaks down.
 The  generalization of (\ref{Redfieldeq})  to include non-Markovian effects
should be considered. 
%

\subsection{Relaxation and dephasing}

Due to the environmental influence, the oscillatory motion of the undamped 
system gets damped. Moreover, a frequency shift of the bare oscillation
 frequencies $\omega_{nm}$ occurs. We evaluate  the dephasing rates $\Gamma_{nm}$  and the decoherence rate $\Gamma$ to lowest order in
 the coupling to the bath from the Redfield equations (\ref{Redfieldeq}).
 We find  the expressions $(i={\rm I,II})$
\begin{eqnarray}
\Gamma^{ (i)}&=&-\sum_n {\mathbf R}_{n,n,n,n}({\mathbf \Lambda}^i)
=2\sum_{n>m=1}^4
\gamma^{ i}_{nm}{ S}_i(\omega_{nm})\;, \label{relaxation}
\end{eqnarray} with 
 $ { S}_i(\omega_{nm}):=[J_i(\omega_{nm})/\hbar]\coth(\hbar\omega_{nm}/2k_{\rm B}T)$, and $\gamma^{\rm I}_{nm}:=s^2_{nm}+r^2_{nm}$,
$\gamma^{\rm II}_{nm}:=s^2_{nm}$. Likeways for
nondegenerate levels $(|\omega_{mn}|>|R_{n,m,n,m}|)$ the dephasing rates read 
\begin{eqnarray}
\Gamma_{nm}^{(i)}&=&-{\rm Re}{\mathbf R}_{n,m,n,m}({\mathbf \Lambda}^i)
= \frac{1}{\hbar^2}(s_{nn}-s_{mm})^2 \hat M_i(0) +
\gamma^i_{nm} { S}_i (\omega_{nm})\nonumber\\
\!\!\!\!\!&+& \frac{1}{\hbar^2}
\sum_{k \neq n,m}[\gamma^i_{nk}{\rm Re}\hat M_i(\omega_{kn})+
\gamma^i_{mk}{\rm Re}\hat M_i(\omega_{km})]\;.\nonumber\\
\label{dephasing}
\end{eqnarray}
Thus, the real part  of the Redfield tensor provides the relaxation and dephasing rates. The imaginary part is responsible for an environment induced shift of the  oscillation  frequency $\omega_{mn}$.
One finds $\omega_{nm}\to \tilde\omega_{mn}:= \omega_{nm}-{\rm Im}
{\mathbf R}_{n,m,n,m}$,
where
\begin{eqnarray}
\!\!\!\!-{\rm Im}{\mathbf R}_{n,m,n,m}&=&
(s^2_{nn}-s_{mm}^2){\rm Im}\hat M_i(0)
+ \gamma^i_{nm}\frac{ \omega_{mn}}{\pi\hbar}\int_0^{\omega_c}d\omega
\frac{J_i(\omega)}{\omega^2-\omega_{mn}^2}
\coth\left(\frac{\hbar\omega}{2k_{\rm B}T}\right)
\nonumber \\ 
&+&\frac{1}{\hbar^2}{\rm Im}\sum_{k\neq m,n}[\gamma^i_{nk}\hat M_i(\omega_{kn})+\gamma_{mk}^i
\hat M_i^*(\omega_{km})]\;.\nonumber \\
\end{eqnarray}

We observe         that at zero temperature   the general relation holds 
\begin{equation}
\Gamma^{(i)}= \frac{2}{N_{lev}-1}\sum_{n> m  } \Gamma_{nm}^{(i)}\;, \qquad T=0\;,
\label{2/3}\end{equation}
where $N_{lev}$ is the number of levels involved in the dynamics
 ($N_{lev}=4$ for coupled qubits, and $N_{lev}=2$ for uncoupled qubits). 
It follows that 
the dephasing rates are always smaller than the relaxation rates $\Gamma^{(i)}$.  
It is now interesting to compare 
 (\ref{relaxation}) and (\ref{dephasing}) to  the decay rates 
$\Gamma_{\rm qb}$ and $\Gamma_\phi$ describing relaxation and dephasing,
 respectively,  of the  uncoupled  single qubits. 
An analysis similar to the one here reported yields for the single 
qubit the results \cite{Weiss99}
\begin{eqnarray}
\Gamma_{\rm qb}=\frac{ E_J^2}{2 E^2_{\rm qb}}{ S}(E_{\rm qb})\;,\qquad 
\Gamma_{\phi}=\frac{\Gamma_{\rm qb}}{2} + \frac{ E_z^2}{2E_{\rm qb}^2}
{S}(0)\;,\label{rates1qb}
\end{eqnarray}
where ${ S}(\omega)=[J_V(\omega)/\hbar]\coth(\hbar\omega/2k_{\rm B}T)$.  
As shown by (\ref{relaxation}),
incoherent processes contributing to $\Gamma^{(i)}$
increase  with the number of levels involved. Thus, 
 the coupled qubits relaxation rates $\Gamma^{(i)}$ 
are larger than  the single qubit relaxation rate $\Gamma_{\rm qb}$,
 cf. also next Section and Fig. 4. As far as regards the dephasing rates,                 
    only the last term in (\ref{dephasing}) is not present for a single qubit.
 This term, however,  is expected to 
 play  a minor role. Hence,  
  the dephasing rates $\Gamma_{m,n}$ remain of the same order of the
single qubit relaxation rate $\Gamma_\phi$, cf. Fig. 4.


\section{Dynamics of dissipative coupled charge qubits} 

By gathering all of the results of the previous Sections 2 and 3, 
 we evaluate now the 
occupation probabilities $P_\mu(t):= \rho_{\mu\mu}(t)$, with 
 the vector  $|\mu\rangle
 \in \{|\uparrow \uparrow \rangle, |\downarrow\downarrow\rangle,
 (|\uparrow\downarrow\rangle \pm |\downarrow \uparrow \rangle)/{\sqrt 2}  \}$. 
Given an initial reduced density matrix $\rho (t_0)$, we obtain
 with $t_0=0$ and to linear order in the coupling parameter $\alpha$, 
 the result 
\begin{eqnarray}
P_\mu(t)&=&\sum_n a^2_{\mu n}[\rho_{nn}(t_0)-\rho_{nn}^\infty] e^{-\Gamma t} + P_\mu^\infty\nonumber\\
 \!\!\!\!\!\!&+&
\sum_{n\neq m = 1}^4 a_{\mu n}a_{\mu m}
 [ 
\rho_{nm}(t_0)-\rho_{nm}^\infty]e^{(i\tilde\omega_{nm}-\Gamma_{nm})t},
\nonumber\\
\label{anal}
\end{eqnarray}
where $a_{\mu n}:=\langle \mu|m\rangle$ are the matrix elements for the
transformation from the eiegenvectors basis to the singlet/triplet representation,
cf. (\ref{transform}). Moreover, the stationary equilibrium value is
\begin{equation}
P_\mu^\infty=\sum_{n,m} a_{\mu n}a_{\mu m}\rho_{nm}^\infty\;.
\end{equation}
Note that, due to dissipation, the offdiagonal elements $\rho_{nm}^\infty:=\rho_{mn}(t=\infty)$, 
 $n\neq m$, are different 
from zero. They are terms of   order $\alpha$ and vanish with $\alpha\to 0$.

In the following we present results of a numerical integration of 
the Redfield equations (\ref{Redfieldeq}). We consider   the 
case   of  identical qubits coupled to the  same bath with spectral density   $J_V(\omega)=2\pi\hbar\alpha\omega$. 
For the discussion of the results we find it convenient to introduce the 
frequency $\omega_{\rm J}:=E_{\rm J}/\hbar$, where $E_{\rm J}=E_{\rm J}^1=
 E_{\rm J}^2.$ We start by showing in Fig. 3 a comparison between 
 the analytical prediction in Eq. (\ref{anal}) and results
 of the numerical evaluation of the Redfield equations. 
  In particular, we plot  
in Fig. 3 the time evolution of the survival probability $P_{\uparrow\uparrow}(t)$ of finding the system in the triplet state 
$|\uparrow\uparrow\rangle$ in which it was prepared at the initial time $t_0=0$. The different oscillatory bits corresponding to the frequencies $\tilde \omega_{nm}$ are clearly seen. A good agreement with the analytical solution Eq. (\ref{anal}) is found. We observe that, since in (\ref{anal}) only linear terms
in $\alpha$ are kept, deviations from the numerical solution increase with increasing of the coupling parameter $\alpha$ (not shown).  
 We also note that, to check
 the validity of the Markovian approximation involved in the derivation of
 the Eqs. (\ref{Redfieldeq}), a numerical integration which abandoned the 
 Markov assumption was performed. In the chosen parameter regime a perfect
agreement with the Markovian-Redfield result was found (not shown).
Figs. 4 shows results for the survival 
 probability   $P_{\uparrow\uparrow}(t)$, and for the occupation
probability      $P_{\rm ent}(t)$ of the entangled state $(|\uparrow\downarrow
\rangle+|\downarrow\uparrow\rangle)/
\sqrt{2}$ 
for two different values of $\alpha$. 
The initial condition 
 $\rho(t_0)=|\uparrow \uparrow\rangle$ is assumed. The figure clearly shows a 
strong damping of the coherent motion already for the ``small'' damping 
constant value $\alpha=0.01$. However, when $\alpha=0.001$ coherence is preserved over many coherent oscillation periods. Note also the phase shift induced 
by  the different choice of the damping parameter $\alpha$.   
Fig. 5  shows the relaxation  rate $\Gamma^{}$ (dashed line)
 and dephasing rate $\Gamma_{14}^{}$ (solid line),  versus temperature.
 For comparison also the 
 $\Gamma_\phi$ (\ref{rates1qb})  relative to a single qubit is shown. Clearly, the single qubit rates turns out to be smaller than its coupled qubit counterpart, but it is still of the same order of magnitude.  
When the qubit relaxation  rate  $\Gamma_{\rm qb}$ is compared to the coupled qubits decay rate $\Gamma$, however, a larger increase of $\Gamma$ in comparison
to $\Gamma_{\rm qb}$ is found, as expected from (\ref{relaxation}) (not shown).

Finally, in Fig. 6 results for the relaxation and dephasing rates  versus the
coupling parameter $-\gamma$ are shown. To better outline the differences 
with respect to the single qubit case, the rates have been normalized to the
 dephasing rate $\Gamma_\phi$ of a single qubit. Note that  in the 
absence of coupling is $\Gamma_{14}=\Gamma_{13}=2\Gamma_\phi$. Moreover,
the relation (\ref{2/3}) always holds. 


\section{Conclusions and outlook}

In conclusion, we evaluated the decoherence and dephasing rates of coupled qubits for two
possible dissipative configurations. The first case, denoted I, dealt with
the case in which the coupled qubits experience  the same dissipative 
 forces, arising from a common electromagnetic environment.
In the second case, denoted II, each qubit is coupled to its own oscillator
bath.  Due to enhanced correlations, dissipative effects are stronger in configuration I. 
Interestingly enough, we found that in the degenerate case, in which the
two qubits are characterized by the same Josephson coupling energy and
 the same asymmetry energy, the nondissipative Hamiltonian becomes separated 
into two blocks, corresponding to  triplet and singlet states, respectively.
Hence, a system prepared in one of the triplet states undergoes quantum coherent oscillations among the three states, while no motion is associated to the
 singlet state. When the coupling to the correlated environment I is 
included, thermalization of the triplet states occurs, while the singlet
state remains unaffected. In contrast, for the case of uncorrelated baths
 II, environmental induced transitions between the singlet and triplet 
states occurs. Hence, also the singlet state thermalizes. 
  
When compared with the relaxation and dephasing rates of a single
qubit, it turns out that the coupled qubits relaxation rate scales with
the number of allowed interlevel transitions.  
This yields an enhancement of decoherence effects as compared to the single qubit case. We performed explicit calculations which kept into account the
effects of fluctuation of the voltage sources. In fact, this is thought
to be the dominant decoherence mechanism in Josephson qubits.
 However, 
the effect of other noise sources can be  esaily icluded in our formalism.  
\\

\section*{Acknowledgements}

We thank Yu. Makhlin for stimulating discussions. 
This work has been supported by the programme
`Quanten-Informationsverarbeitung' of the
German Science Foundation (DFG), the European TMR network 
`Superconducting Nano-Circuits' and the IST programme `SQUBIT'.

\begin{figure}[t,h]
\begin{center}
\includegraphics*[width=5cm]{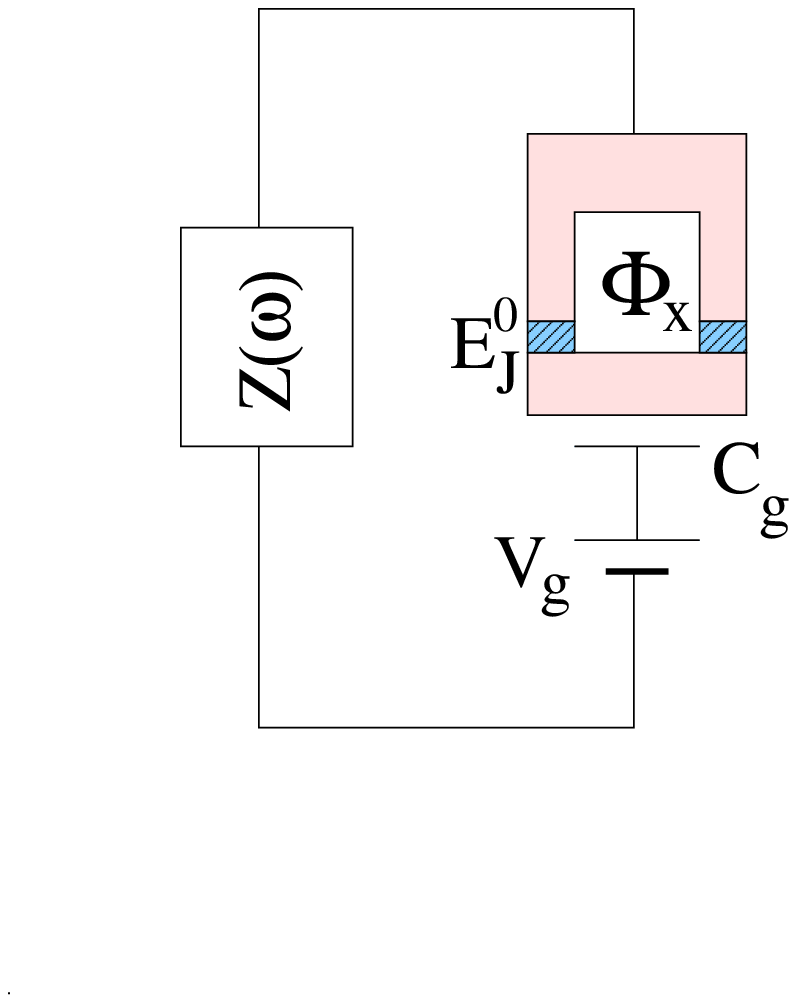}
\caption{Charge  qubit with  Josephson coupling controlled by an external field
$\Phi_X$ threading the dc-SQUID. The dissipative influence originating by voltage fluctuations are captured by an impedence $Z(\omega)$.  
 \label{fig1 }}
\end{center}
\end{figure}

\begin{figure}[t,h]
\includegraphics*[width=5cm]{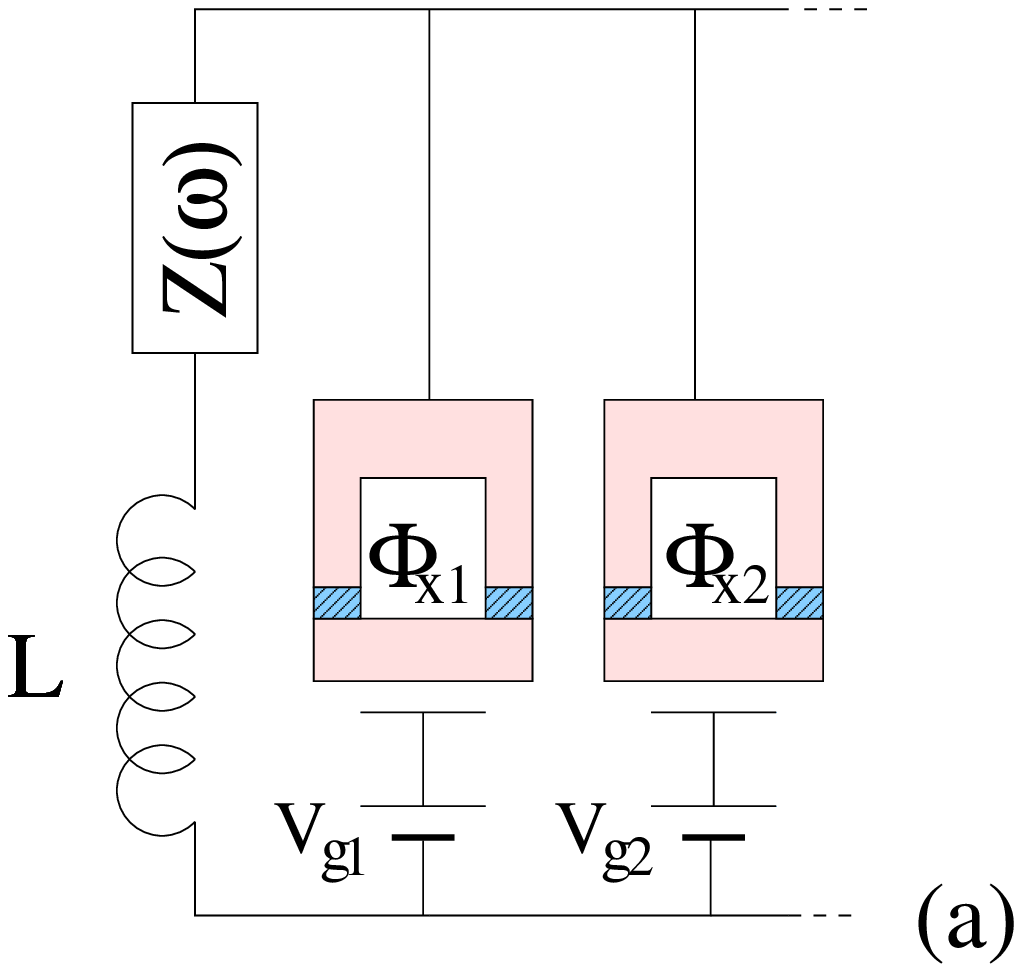}
\includegraphics*[width=5cm]{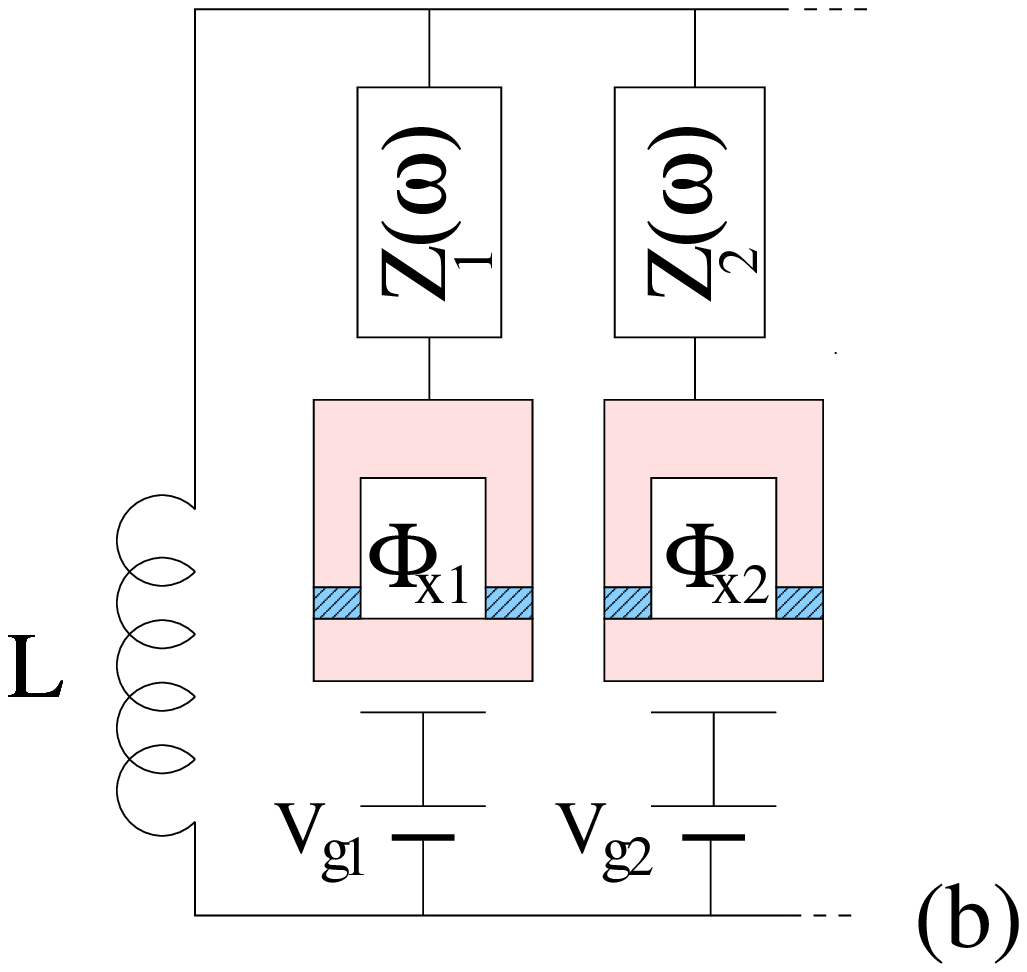}
\caption{A register of many charge qubits coupled to a common inductor $L$.
In case a) the qubits are subject to a common dissipative environment,
while in configuration b) each qubit is coupled to its own oscillator bath.
\label{fig2b }}
\end{figure}

\begin{figure}[t,h]
\begin{center}
\includegraphics*[width=7cm]{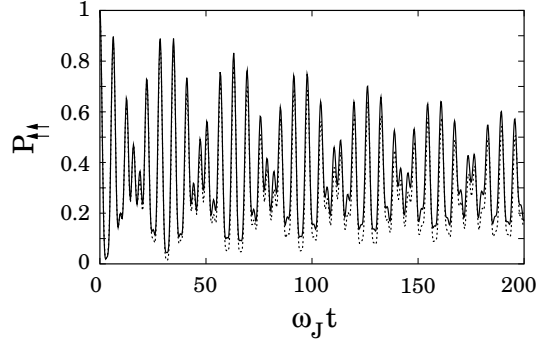}
\caption{Time evolution of the survival probability $P_{\uparrow\uparrow}$
 for coupled qubits being prepared in the state $|\uparrow\uparrow\rangle$.
 The solid line denote results of a numerical integration of the
Redfield Eqs. (\ref{Redfieldeq}), while the dotted line refer to the
analytical solution (\ref{anal}). We choose Ohmic constant $\alpha=0.0001$
and zero temperature.  Here and in the next figures we introduced 
$\omega_{\rm J}=E_{\rm J}/\hbar$ and we set $
 E_z^1=E_z^2=0,
\gamma=-0.1E_{\rm J}, \omega_c =50\omega_{\rm J}$. 
 \label{fig3 }}
\end{center}
\end{figure}

\begin{figure}[t,h]
\begin{center}
\includegraphics*[width=6cm]{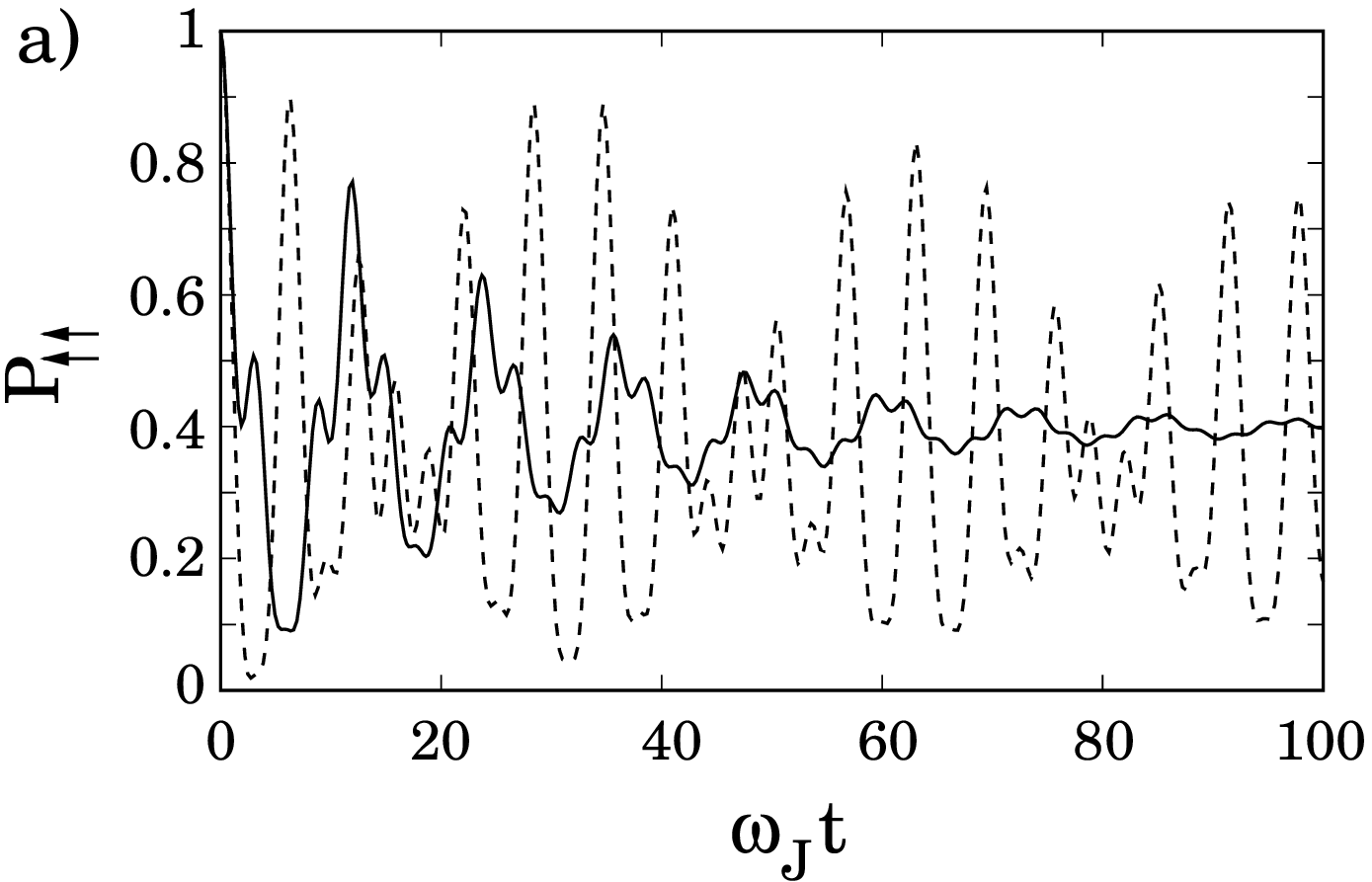}
\includegraphics*[width=6cm]{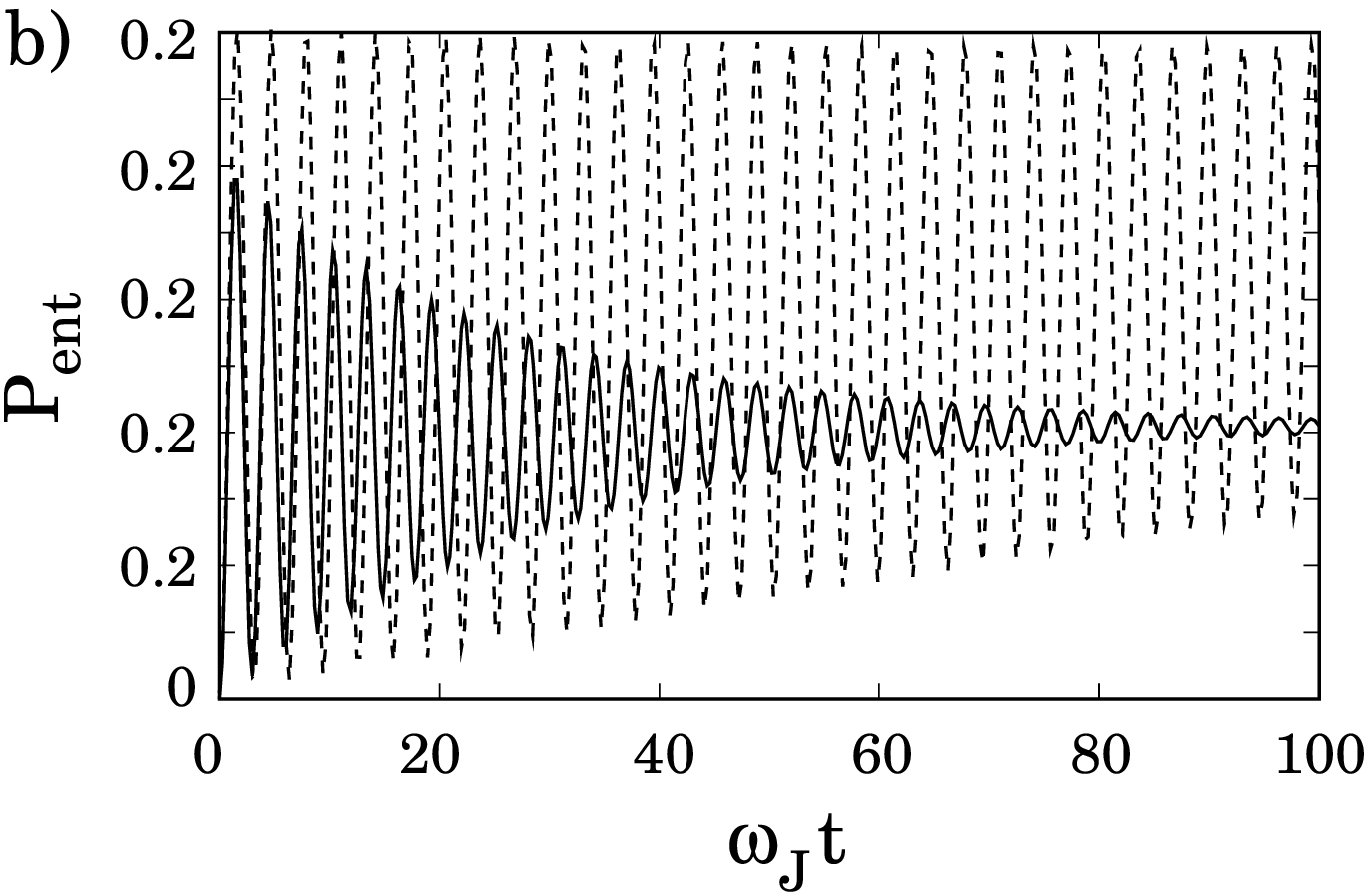}
\caption{a) Time evolution of the survival probability 
 $P_{\uparrow\uparrow}$, and b) of the probability of the triplet entagled state $P_{\rm ent}(t)$
 for coupled qubits being prepared in the state $|\uparrow\uparrow\rangle$.
 The  curves, obtained from a numerical integration of the
Redfield Eqs. (\ref{Redfieldeq}), show results at $T=0$ for coupling parameter 
$\alpha=0.001$ (dashed line), and $\alpha=0.01$ (solid line). 
 \label{fig4 }}
\end{center}
\end{figure}

\begin{figure}[t,h]
\begin{center}
\includegraphics*[width=7.3cm]{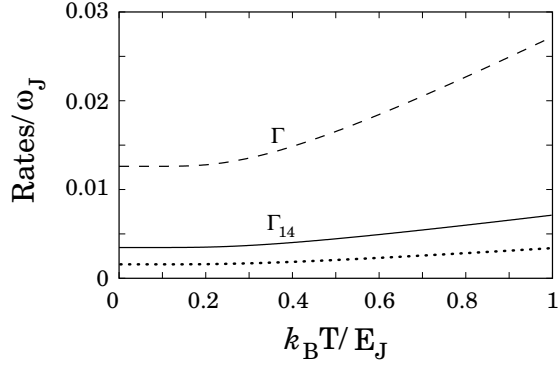}
\caption{Relaxation rate $\Gamma$ and dephasing rate $\Gamma_{14}$ of
 coupled qubits versus
 temperature. For comparison also the single qubit dephasing rate 
 $\Gamma_\phi$ is shown (dotted line).
 \label{fig5 }}
\end{center}
\end{figure}

\begin{figure}[t,h]
\begin{center}
\includegraphics*[width=7cm]{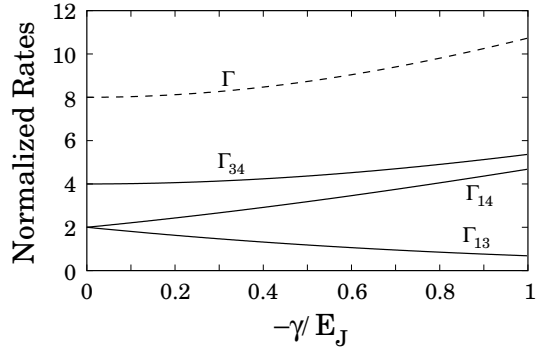}
\caption{ Relaxation rate $\Gamma$ and dephasing rates $\Gamma_{34},
 \Gamma_{13},$ and $\Gamma_{14}$ of
 coupled qubits versus
 the qubits coupling parameter $-\gamma$. The rates are normalized to the
 single qubit dephasing rate $\Gamma_\phi$. 
 The  temperature  is set to zero. 
 \label{fig6 }}
\end{center}
\end{figure}

\end{document}